\documentclass[iop]{emulateapj}
\usepackage{epsfig}
\usepackage{apjfonts}
\usepackage{aas_macros}

\begin{document}

\title{Viewing angle constraints on  S190425z and S190426c and the joint gravitational-wave/gamma-ray detection fractions for binary neutron star mergers}
\author{Hao-Ran Song$^{1}$, Shun-Ke Ai$^{1}$, Min-Hao Wang$^{1}$, Nan Xing$^{1}$, He Gao$^{1,*}$, and Bing Zhang$^{2}$}
\affiliation{
$^1$Department of Astronomy, Beijing Normal University, Beijing 100875, China; gaohe@bnu.edu.cn\\$^2$Department of Physics and Astronomy, University of Nevada Las Vegas, NV 89154, USA.
  }

\begin{abstract}
The LIGO and Virgo scientific collaboration (LVC) alerted two binary neutron star (BNS) merger candidates, S190425z and S190426c. Fermi-GBM observed 55.6\% (for S190425z) and 100\% (for S190426c) of the probability regions of both events at the respective merger times, but no gamma-ray burst (GRB) was detected in either case. The derived luminosity upper limits suggest that a short GRB similar to GRB 170817A would not be detectable for both cases due to their larger distances than GW170817. Assuming that the jet profile obtained from the GW170817/GRB 170817A is quasi-universal for all BNS-GRB associations, we derive that the viewing angles of S190425z and S190426c should be $> (0.11-0.41)$ and $> (0.09-0.39)$, respectively. Through Monte Carlo simulations, we show that with the GRB 170817A-like jet structure, all sky gamma-ray detectors, such as GBM and GECAM, are expected to detect $\sim 4.6\%$, $3.9\%$, $1.7\%$ and  $6.6\%$, $5.7\%$, $2.8\%$ BNS mergers triggered by aLIGO, aLIGO A+ and ET, respectively. The joint detection fraction would be largely reduced for Swift-BAT, SVOM-ECLAIRS and Einstein Probe, whose sensitivities are better but fields of view are smaller.
\end{abstract}

\keywords{gamma-ray burst: general - gravitational waves}


\section{INTRODUCTION}

After the exciting discovery of the first binary neutron star (BNS) merger event GW170817 \citep{abbott17a} and its associated gamma-ray burst (GRB) 170817A, kilonova, and multi-wavelength afterglows \citep{abbott17b}, the LIGO-Virgo scientific collaboration (LVC) lately reported two more candidates that may have the BNS merger origin, i.e., LIGO/Virgo S190425z and LIGO/Virgo S190426c \citep{LVC25z,LVC26c}. 

S190425z was identified during the real-time processing of data from LIGO Livingston Observatory (L1) and the Virgo Observatory (V1) at 2019-04-25 08:18:05.017 UTC \citep{LVC25z}. The false alarm rate is estimated by the online analysis as $4.5\times10^{-13}$ Hz, or about one in 70 thousands of years. Since the source was not detected by LIGO Hanford (H1) and the signal-to-noise ratio (SNR) was below the threshold in V1, LVC provides a very poor localization constraint. Assuming that the candidate is astrophysical in origin, the probability for classifying this GW event as BNS merger is greater than $99\%$. Detailed data analysis is on going. Multi-band observations from radio to gamma-rays were immediately operated to search for its electromagnetic (EM) counterpart candidate, but no confident counterpart was identified. The Gamma-ray Burst Monitor (GBM) on board the Fermi Gamma-Ray Observatory (Fermi-GBM) observed 55.6\% of the probability region at the merger event time \citep{GBM25z}. There was no onboard trigger around the event time, and no counterpart candidate was identified with both automated, blind search and coherent search for a GRB signal (from $\pm$30 s around the merger time). The Fermi-GBM Team thus estimated the intrinsic luminosity upper limit for a S190425z-associated-GRB, if any, as $(0.03-6.2) \times 10^{49} {\rm erg~s^{-1}}$. 

S190426c was identified during the real-time processing of data from LIGO Hanford Observatory (H1), LIGO Livingston Observatory (L1) and Virgo Observatory (V1) at 2019-04-26 15:21:55.337 UTC \citep{LVC26c}. The false alarm rate for this event is estimated by the online analysis as $1.9\times10^{-8}$ Hz, or about one in 1 year and 7 months. Assuming that the candidate is of an astrophysical in origin, the probability for classifying this GW event as a BNS merger is $49\%$. Detailed data analysis is on going. Similar to S190425z, multi-band observations were carried out immediately to search for an EM counterpart of S190426c, but no confident counterpart has been identified yet. For S190426c, Fermi-GBM was observing 100\% of the probability region at the merger event time \citep{GBM26c}. Again, there was no onboard trigger around the event time, and no counterpart candidate was identified with both automated, blind search and coherent search (from $\pm$30 s around the merger time). The Fermi-GBM Team thus estimated the intrinsic luminosity upper limit for a S190426c-associated-GRB, if any, as $(0.09-10.6) \times 10^{49} {\rm erg~s^{-1}}$. 

In this work, we show that the luminosity upper limit in the $\gamma$-ray band could lead to interesting constraint on the viewing angles of S190425z and S190426c, provided that S190425z and S190426c are associated with a short GRB, whose jet profile is similar to that of GRB 170817A. Moreover, under the hypothesis that the jet profile obtained from the GW170817/GRB 170817A is quasi-universal for all BNS-GRB associations, we perform Monte Carlo simulations to estimate what fraction of BNS mergers detectable by GW detectors is expected to be simultaneously detected in $\gamma$-rays.

\section{Constraints on the viewing angle for S190425z and S190426c}
\begin{figure*}[ht!]
\centering
\resizebox{90mm}{!}{\includegraphics[]{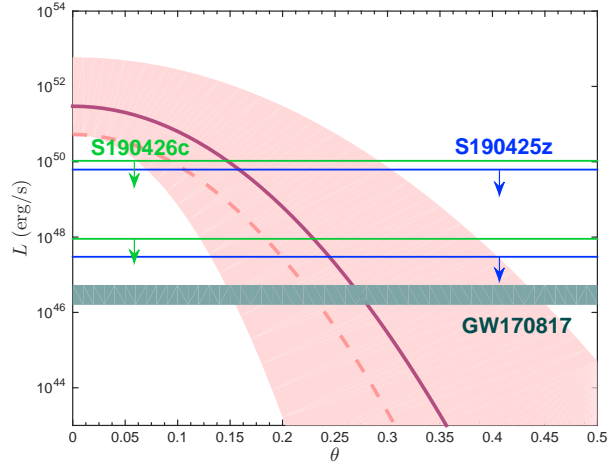}}
\caption{The purple solid line and pink filled region present the jet profile of GRB 170817A and its 1-$\sigma$ region based on \cite{troja18}. The dark green region presents the $\gamma$-ray emission luminosity of GRB 170817A and the light blue and green solid lines show $\gamma$-ray emission luminosity upper limits for S190425z and S190426c. The dashed pink line presents the lower boundary of GRB 170817A jet profile proposed in \cite{finstad18}.}
\label{constraint}
\end{figure*}

The abnormally low prompt emission luminosity \citep{goldstein17,zhang18} and the slow rising of the multi-wavelength lightcurves \citep{troja17,mooley18,lazzati18,troja18,lyman18} of GRB 170817A  suggested that the event is best interpreted as a structured jet \citep[e.g.][]{zhang02,rossi02} with a large viewing angle from the jet axis. With a broad-band study and a multi-messenger analysis including the GW constraints, {it has been proposed that the data of GW170817/GRB 170817A favor a Gaussian-shaped jet profile  \citep{zhang02,troja18,alexander18,mooley18,lazzati18,ghirlanda19}}
\begin{eqnarray}
E(\theta)=E_0{\rm exp\left(-{\theta^2 \over 2\theta^2_c}\right)}
\end{eqnarray}
for $\theta \leq \theta_w$, where $E_0$ is the on-axis equivalent isotropic energy, $\theta_c$ is the charateristic angle of the core, and $\theta_w$ is the truncating angle of the jet. Such a Gaussian jet profile seems to be supported by numerical simulations of a short GRB jet propagates in a dynamical ejecta with a negligible waiting time of jet launching \citep{xie18,geng19}.
In order to interpret the multi-wavelength EM observations and the viewing angle constraint from the GW analysis, \cite{troja18} proposed $\theta_c=0.057_{-0.023}^{+0.025}$, ${\rm log_{10}}E_0=52.73_{-0.75}^{+1.3}$ and $\theta_w=0.62_{-0.37}^{+0.65}$ for the jet profile of GRB 170817A. The value of the Hubble constant reported by the Planck collaboration \citep{planck} was adopted.

In Figure 1, we plot the jet profile of GRB 170817A with the 1-$\sigma$ region as proposed by \cite{troja18}, the $\gamma$-ray emission luminosity of GRB 170817A, and the $\gamma$-ray emission luminosity upper limits for S190425z and S190426c. Note that in order to convert the energy profile in \cite{troja18} to the luminosity profile, here we assume that the $\gamma$-ray radiation efficiency is 10\%, the burst duration $T_{90} \sim 2$ s and the spectrum is flat. {Such a jet profile covers the regime of known short GRBs, the hypothesis of a quasi-universal structured jet for all short GRBs seems to be consistent with the currently available data \citep{beniamini19a,salafia19}}. Assuming that both S190425z and S190426c are associated with a short GRB, whose jet profile is similar to that of GRB 170817A, we derive that the viewing angle of S190426c is $> (0.06-0.39)$, with the uncertainty mainly defined by the uncertainty of its luminosity distance. {As shown in Figure 1, the uncertainty of the jet profile considered here is already quite wide to accommodate a variety of possible quasi-universal structures, so our conclusion is generally valid if the jet structure of individual GRBs deviates from the Gaussian form but generally falls into the shaded region in Figure \ref{constraint}.} For S190425z, with an additional assumption that its location was within the filed of view of Fermi-GBM, one can derive its viewing angle being $> (0.07-0.41)$, with the uncertainty again mainly defined by the uncertainty of its luminosity distance at the merger time. It is worth noticing that \cite{finstad18} performed a joint analysis of the GW/EM observations and suggested a conservative lower limit on the viewing angle of $>13^{\circ}$ for GRB 170817A. If one takes this limit, the lower boundary of GRB 170817A jet profile would become tighter (dashed line in Fig.1), so that the viewing angle constraint for both S190425z and S190426c would become tighter, i.e., $> (0.09-0.39)$ for S190426c and $> (0.11-0.41)$ for S190425z, respectively. 

\section{$\gamma$-ray detection rate for GW triggers}

Assuming that the jet profile obtained from the GW170817/GRB 170817A is quasi-universal for all BNS-GRB associations, here we perform Monte Carlo simulations to estimate what fraction of BNS mergers detectable by GW detectors is expected to be also detected by $\gamma$-ray detectors, such as Fermi-GBM \citep{meegan09}, Neil Gehrels Swift-BAT \citep{gehrels04}, the future Chinese-French GRB mission SVOM-ECLAIRS \citep{gotz14}, and the future Chinese mission Gravitational wave high-energy Electromagnetic Counterpart All-sky Monitor \cite[GECAM;][]{gecam}, and by large FoV X-ray detectors, such as the future Chinese mission Einstein Probe \cite[EP;][]{yuan18}.
\begin{figure*}  
\centering  
\includegraphics[height=9cm,width=16cm]{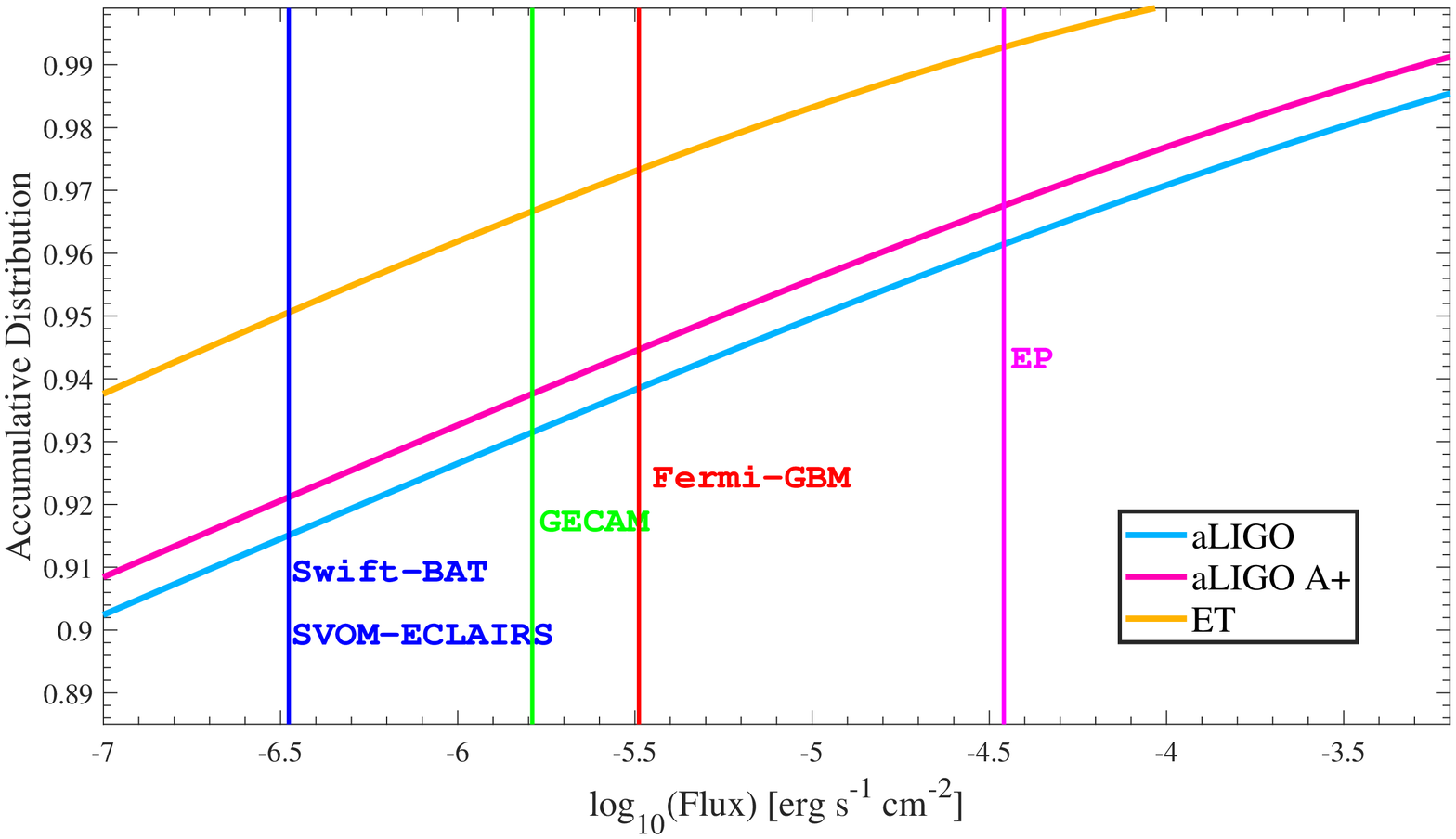}
\caption{The accumulative distribution of bolometric flux in the $1-10^4$ keV band for BNS mergers triggered by aLIGO, aLIGO A+ and ET. The vertical colored lines show the effective sensitivity of Fermi-GBM,  Swift-BAT, SVOM-ECLAIRS, GECAM and EP, which correspond to Band function spectrum with $\alpha=-1$, $\beta=-2.3$ and $E_{p}$ equaling to the mean value for the simulated sample.}
\end{figure*} 

\begin{table*}
\caption{The joint detection rate of five gamma-ray missions and three GW detectors.}
\centering
\begin{tabular}{cccccc}            
\hline\hline
            &      Fermi-GBM         &       GECAM          &          Swift-BAT              &             SVOM-ECLAIRS               & EP   \\
\hline
aLIGO       &  4.61\%\ (1.83/yr)      &   6.57\%\ (2.61/yr)   &         0.977\%\ (0.388/yr)            &       1.68\%\ (0.668/yr)   &    0.366\%\ (0.145/yr)       \\
\hline
aLIGO A+    &  3.87\%\ (8.40/yr)      &   5.72\%\ (12.4/yr)   &         0.887\%\ (1.92/yr)            &        1.53\%\ (3.31/yr)       &   0.266\%\ (0.578/yr)    \\
\hline
ET          &  1.69\%\ (593/yr)      &   2.83\%\ (990/yr)   &          0.536\%\ (188/yr)            &        0.924\%\ (324/yr)     &   0.112\%\ (39.2/yr)      \\
\hline\hline
\end{tabular}
\label{table1}
\end{table*}

\subsection{Sample for BNS mergers detectable by GW detectors}

The event rate density for BNS mergers at a given redshift $z$ could be estimated as 
\begin{eqnarray}
N(z)={R_{\rm NS-NS~merger,0}\times f(z)\over 1+z}{dV(z) \over dz},
\label{eq:Nz}
\end{eqnarray}
where $R_{\rm NS-NS~merger,0}=1540_{-1220}^{+3200} {\rm Gpc^{-3}yr^{-1}}$ is the local NS-NS merger rate  \citep{abbott17a}, $f(z)$ is the the dimensionless redshift distribution factor and $dV(z)/dz$ is the differential comoving volume $dV(z)/dz$, which reads
\begin{eqnarray}
{dV(z) \over dz}&=&4\pi\left({c \over H_0}\right)^3 \left(\int_{0}^z {dz \over \sqrt{1-\Omega_m+\Omega_m(1+z)^3}}\right)^2 \nonumber \\
&\times&{1 \over \sqrt{1-\Omega_m+\Omega_m(1+z)^3}},
\end{eqnarray}
Here Planck results are adopted for cosmological parameters, e.g. $H_0=67.8{\rm km~s^{-1}Mpc^{-1}}$, $\Omega_m=0.308$ and $\Omega_\Lambda=0.692$ \citep{planck}. 

The function $f(z)$ depends on the initial redshift distribution of BNS systems and the delay time distribution for BNS systems going through the inspiral phase to the merger. We first assume the initial redshift distribution of BNS systems track the Star formation rate (SFR) in units of $M_{\odot}{\rm Gpc^{-3}yr^{-1}}$, which could be empirically expressed as
\citep{yuksel08}:
\begin{eqnarray}
{\rm SFR(z)}\propto\left[\left(1+z\right)^{3.4\eta}+\left({1+z \over 5000}\right)^{-0.3\eta}+\left({1+z \over 9}\right)^{-3.5\eta}\right]^{1 \over \eta}.
\label{eq:SFR}
\end{eqnarray}
\cite{sun15} have investigated three main types of delay time distributions (i.e. Gaussian distribution \citep{virgili11}, power-law distribution \citep{piran92} and log-normal distribution \citep{wanderman15}) and suggested that the power-law model leads to a wider redshift distribution of NS-NS merger than other two models, which is disfavored by the sGRB data. In this work, log-normal delay time distribution is assumed as an example, whose formula reads
\begin{eqnarray} 
m(\tau)={\rm exp}\left(-{({\rm ln}\tau-{\rm ln}t_d)^2 \over 2\sigma^2}\right)/(\sqrt{2\pi}\sigma),
\end{eqnarray}
with $t_d=2.9~{\rm Gyr}$ and $\sigma=0.2$. We randomly generate $10^7$ binary NS systems following the redshift distribution described by equation \ref{eq:Nz}, with redshift ranging from 0 to 3. 

For each binary NS system, their NS masses ($m_1$ and $m_2$) are generated following the observationally-derived distribution of Galactic NS-NS systems, i.e. $M_{\rm BNS}$ has a normal distribution $N(\mu_{\rm BNS}=1.32\,M_{\odot},\sigma_{\rm BNS}=0.11)$,
with a mean $\mu_{\rm BNS}$ and a standard deviation $\sigma_{\rm BNS}$ \citep{kiziltan13}, and their viewing angles are generated with a uniform distribution in the directional space ($N(\theta)\propto \sin (\theta)$). 

Here we estimate the GW signal's amplitude during the inspiral phase within the Newtonian approximation condition as
\begin{eqnarray}
h_{+}=-{2G\mu \over c^2D_L}(1+{\rm cos}^2 \theta)\left({v \over c}\right)^2{\rm cos}2\varphi
\end{eqnarray}
\begin{eqnarray}
h_{\times}=-{4G\mu \over c^2D_L}{\rm cos} \theta \left({v \over c}\right)^2{\rm sin}2\varphi,
\end{eqnarray}
where $+$ and $\times$ are two polarizations of the GW signal, $\mu=m_1 m_2/(m_1+m_2)$ is the reduced mass, $D_L$ presents the luminosity distance of the BNS system, $\theta$ is the viewing angle, $v=(\pi G M f)^{1/3}$ is the Equivalent linear velocity ($f=2f_{\rm orbital}=\omega/\pi$ is GW signal's frequency). The evolution of $v$ over time is
\begin{eqnarray}
{d(v/c) \over dt}={32 \eta \over 5}{c^3 \over GM}\left({\nu \over c}\right)^9,
\end{eqnarray} 
where $M=m_1+m_2$ is the total mass and $\eta=(m_1 m_2)/M^{2}$ is the dimensionless reduced mass. 

Given the sensitivity curve of GW detectors, the signal to noise (SNR) of GW detection for each BNS mergers could be estimated as \citep{corsi09}
\begin{eqnarray}
\left <{\rm SNR} \right >^2= \int_{f_{\rm min}}^{f_{\rm max}} {h_c(f)^2 \over fS_n} d \ln f.
\end{eqnarray}
	where $S_n$ is the power spectral density (PSD) of the detector noise, $f_{\rm min}$ and $f_{\rm max}$ define the frequency range of the signal, from 10 Hz to 2000 Hz, $h_c$ is a characteristic amplitude in the frequency domain, which reads 
\begin{eqnarray}
h_c=fh(t)\sqrt{ dt \over df}.
\end{eqnarray}
where $h(t)=\sqrt{h_{+}^2+h_{\times}^2}$. Here we consider the designed PSD for aLIGO \citep{abramovici92} and aLIGO A+ \citep{ligo16}, as well as the proposed third generation GW detectors (i.e. ET \citep{et17}, CE \citep{abbott17}). We set SNR larger than 16 as the criteria for GW triggers in our simulation. 

\subsection{$\gamma$-ray flux distribution and detection rate}

We assume that every GW triggered event in our simulation successfully launches a relativistic jet with a quasi-universal jet profile described in Equation 1. For each case, $E_0$, $\theta_c$ and $\theta_w$ are generated following the best-fit distribution inferred from GRB 170817 \citep{troja18}, i.e., $\theta_c=0.057_{-0.023}^{+0.025}$, ${\rm log_{10}}E_0=52.73_{-0.75}^{+1.3}$ and $\theta_w=0.62_{-0.37}^{+0.65}$. Given the viewing angle $\theta$, the $\gamma$-band flux for each BNS system could be estimated as
\begin{eqnarray}
F_{\gamma}=\frac{E_0 \eta_{\gamma}}{4\pi D_L^2}{\rm exp\left(-{\theta^2 \over 2\theta^2_c}\right)},
\end{eqnarray}
where $\eta_{\gamma}$ is the radiative efficiency. Here we adopt $\eta_{\gamma}=0.1$ for the bolometric energy flux in the $1-10^4$ keV band. Assuming that the gamma-ray spectrum for all simulated BNS-associated-GRBs follow the Band function with photon indices $-1$ and $-2.3$ \citep{preece00} below and above $E_p$, respectively, we can estimate the corresponding flux for a specific $\gamma$-ray detector with a given observational energy band, and then justify whether or not the simulated source can be detected. Note that the bolometric isotropic luminosity and the peak energy for GRB 170817A does not satisfy the Yonetoku relation \citep{yonetoku04}. Within the Gaussian structure jet framework, \cite{ioka19} proposed that the profile that $E_p$ changes with the viewing angle $\theta$ should be $E_p(\theta)=E_{p,0}\times \left(1+{\theta / \theta_c}\right)^{-0.4}$, where $E_{p,0}$ and the central luminosity of the Gaussian jet satisfy the Yonetoku relation. Such a prescription  allows to incorporate the GRB 170817A observations with the historical SGRB statistical data. For each simulated GRB, we use such a profile to estimate its $E_p$ value. The accumulative distribution of bolometric $F_{\gamma}$ (in the $1-10^4$ keV band) for different GW detectors are shown in Figure 2, together with the effective sensitivity limit for various $\gamma$-ray detectors. Here the sensitivity for Fermi-GBM is adopted as $\rm \sim2\times 10^{-7}~erg\ s^{-1}$ in $50$ keV to $300$ keV \citep{meegan09}, and the sensitivity for GECAM is adopted as $\rm \sim1\times 10^{-7}~erg\ s^{-1}$ in $50$ keV to $300$ keV \citep{gecam}. For Swift-BAT and SVOM-ECLAIRS, we adopt the same sensitivity as $\rm \sim1.2\times 10^{-8}~erg\ s^{-1}$ in $15$ keV to $150$ keV \citep{gehrels04,gotz14}. The sensitivity for EP is adopted as $\rm \sim3\times 10^{-9}~erg\ s^{-1}$ in $0.5$ keV to $4$ keV \citep{EP}. 

We can see that with GRB 170817A-like jet structure, less than $10\%$ of GW-triggered BNS mergers would have the $\gamma$-ray flux higher than the threshold of the current (or near future) $\gamma$-ray detectors. For instance, the $\gamma$-ray flux of 6.3\% of the aLIGO detectable BNS mergers are above the sensitivity limit of Fermi-GBM. Considering that the average filed of view of Fermi-GBM is around $3/4$ of the whole sky, $\sim$ 4.6\% of BNS mergers detectable by aLIGO is expected to be simultaneously detected by Fermi-GBM. As more distant sources become detectable, the fraction for GW-triggered BNSs being simultaneously detected by Fermi-GBM would be reduced to $\sim$ 3.9\% and $\sim$ 1.7\% for aLIGO A+ and ET, but the absolute detection rate should largely increase. The proposed sensitivity of GECAM is slightly better than Fermi-GBM, and its proposed filed of view is around $4\pi$. In this case, we find that $\sim$ 6.6\%, 5.7\% and 2.8\% of BNS mergers detectable by aLIGO, aLIGO A+ and ET are expected to be simultaneously detected by GECAM, respectively. The sensitivity of Swift-BAT and SVOM-ECLAIRS are better than Fermi-GBM and GECAM, but their fields of view are much smaller [$\sim1/9$ of the whole sky for Swift and $\sim 1/5$ for SVOM-ECLAIRS \citep{chu16}]. In this case, we expect $\sim$ 0.98\%, 0.89\% and 0.54\% of BNS mergers triggered by aLIGO, aLIGO A+ and ET are detectable with Swift-BAT, and $\sim$ 1.7\%, 1.5\% and 0.9\% of BNS mergers triggered by aLIGO, aLIGO A+ and ET are detectable with SVOM-ECLAIRS. The EP has a field of view $\sim 1/11$ of whole sky which results in a small observation rate, $\sim$ 0.37\%, 0.27\% and 0.11\%. The joint detection fraction and absolute joint detection rate of five gamma-ray missions and three GW detectors are collected in Table 1.

\section{CONCLUSION AND DISCUSSION}

Assuming that every BNS merger is associated with a short GRB, whose jet profile is broadly similar to that of GRB 170817A, here we show that even the luminosity upper limit in the $\gamma$-ray band could lead to interesting constraint on the viewing angles of GW triggered BNS mergers. For instance, we derive that the viewing angles of S190425z and S190426c should be $> (0.11-0.41)$ and $> (0.09-0.39)$, respectively. A 170817A-like short GRB would be undetectable for S190425z and/or S190426c due to their larger distances than GW170817/GRB 170817A. The constraints have to be revised in the following situations: 1. S190425z and/or S190426c are not from BNS mergers; 2. S190425z was not in the field of view of GBM (e.g. blocked by Earth); 3. Not all BNS mergers are associated with short GRBs; and 4. BNS short GRBs do not share a quasi-universal jet structure. If our interpretation is correct, however, the constraints on viewing angle would be helpful for GW data analysis to reach better constraints on the binary properties.

Furthermore, with Monte Carlo simulations, we find that with GRB 170817A-like jet structure, all sky gamma-ray detectors, such as GBM and GECAM, are expected to detect $4.6\%$, $3.9\%$, $1.7\%$ and  $6.6\%$, $5.7\%$, $2.8\%$ BNS mergers triggered by aLIGO, aLIGO A+ and ET, respectively. For Swift-BAT, SVOM-ECLAIRS and EP, whose sensitivities are better but fields of view are smaller, the simultaneous GW/$\gamma$-ray detection fraction would be largely reduced. Future joint observations of BNS GW sources by LVC and all sky gamma-ray detectors, such as GBM and GECAM, together with the event rate density studies \citep[e.g.][]{sun15,zhang18}, will finally test the quasi-universal jet hypothesis of BNS mergers.

In our Monte Carlo simulations, in order to justify whether the simulated sources could be detected by the gamma-ray detectors, we simply compare the integrated flux with the effective $\gamma$-ray detector sensitivities. In practice, the detection rates for the various detectors depend on the details of the triggering algorithm and sky background for each detector, particularly for the sources near the effective sensitivity limit. Future work is needed to give more precise predictions for specific $\gamma$-ray detectors, but we expect that the results should be consistent with the results here to orders of magnitude. Also in our simulations we take the posterior jet-parameter distributions from one single measurement of GRB 170817A and regard it as the representative of the physical parameter distributions for all BNS-associated-SGRB populations. The detection rate would be altered if GRB 170817A jet structure is not representative. The detection rate would increase if the jet structure for the population is broader than GRB 170817A, and vice versa. More BNS-associated-SGRB detections in the future will help to better draw the jet structure for the population, which lead to more precise estimations for the detection rates.

\acknowledgments
We thank the referee for the helpful comments, which have helped us to improve the presentation of the paper. This work is supported by the National Natural Science Foundation of China under Grant No. 11690024, 11722324, 11603003, 11633001, the Strategic Priority Research Program of the Chinese Academy of Sciences, Grant No. XDB23040100 and the Fundamental Research Funds for the Central Universities.


\begin{thebibliography}{}
\expandafter\ifx\csname natexlab\endcsname\relax\def\natexlab#1{#1}\fi


\bibitem[Abbott et al.(2017a)]{abbott17a}{Abbott}, B.~P., {Abbott}, R., {Abbott}, T.~D., et al.\ 2017a,PhRvL.,118v1101A

\bibitem[Abbott et al.(2017b)]{abbott17b}{Abbott}, B.~P., {Abbott}, R., {Abbott}, T.~D., et al.\  2017b, \apj, 848L,12A

\bibitem[Abbott et al.(2017c)]{abbott17}{Abbott}, B.~P., {Abbott}, R., {Abbott}, T.~D., et al.\ 2017c, Classical and Quantum Gravit, 34,044001

\bibitem[Abramovici et al.(1992)]{abramovici92} Abramovici, A., Althouse, W.~E., Drever, R.~W.~P., et al.\ 1992, Science, 256, 325

\bibitem[Adhikari et al.(2017)]{ligov17}{Adhikari}, R. X., {Smith}, N., {Brooks}, A., et al. \ 2017, LIGO DCC-T1400226

\bibitem[Alexander et al.(2018)]{alexander18} Alexander, K.~D., Margutti, R., Blanchard, P.~K., et al.\ 2018, \apj, 863, L18

\bibitem[Barthelmy et al.(2005)]{BAT} Barthelmy, S.~D., Barbier, L.~M., Cummings, J.~R., et al.\ 2005, \ssr, 120, 143

\bibitem[Band et al.(1993)]{band93} Band, D., Matteson, J., Ford, L., et al.\ 1993, \apj, 413, 281

\bibitem[Beniamini et al.(2019)]{beniamini19a} Beniamini, P., Petropoulou, M., Barniol Duran, R., et al.\ 2019, \mnras, 483, 840


\bibitem[Chu et al.(2016)]{chu16} Chu, Q., Howell, E.~J., Rowlinson, A., et al.\ 2016, \mnras, 459, 121 

\bibitem[Corsi \& M{\'e}sz{\'a}ros (2009)]{corsi09}{Corsi}, A. \& {M{\'e}sz{\'a}ros}, P., \ 2009, \apj, 702,1171

\bibitem[Fermi/GBM Team (2019a)]{GBM25z} Fermi/GBM Team\ 2019, GRB Coordinates Network, Circular Service, No.~24185

\bibitem[Fermi/GBM Team (2019b)]{GBM26c} Fermi/GBM Team\ 2019, GRB Coordinates Network, Circular Service, No.~24248

\bibitem[Finstad et al.(2018)]{finstad18} Finstad, D., De, S., Brown, D.~A., Berger, E., \& Biwer, C.~M.\ 2018, \apjl, 860, L2 

\bibitem[Geng et al.(2019)]{geng19}{Geng}, J.-J., {Zhang}, B., {K\"olligan}, A. et al.\ 2019, \apjl, 877, L40

\bibitem[Gehrels et al.(2004)]{gehrels04} Gehrels, N., Chincarini, G., Giommi, P., et al.\ 2004, \apj, 611, 1005 

\bibitem[Ghirlanda et al.(2019)]{ghirlanda19} Ghirlanda, G., Salafia, O.~S., Paragi, Z., et al.\ 2019, Science, 363, 968

\bibitem[G{\"o}tz et al.(2014)]{gotz14} G{\"o}tz, D., Osborne, J., Cordier, B., et al.\ 2014, \procspie, 9144, 914423

\bibitem[Goldstein et al. (2017)]{goldstein17} Goldstein, A., Veres P., Burns, E., et al. 2017, ApJL 848, 14L



\bibitem[Ioka \& Nakamura(2019)]{ioka19} Ioka, K., \& Nakamura, T.\ 2019, \mnras, 487, 4884

\bibitem[Kiziltan et al.(2013)]{kiziltan13} Kiziltan, B., Kottas, A., De Yoreo, M., \& Thorsett, S.~E.\ 2013, \apj, 778, 66

\bibitem[Lazzati et al.(2018)]{lazzati18} Lazzati, D., Perna, R., Morsony, B.~J., et al.\ 2018, \prl, 120, 241103

\bibitem[Lyman et al.(2018)]{lyman18} Lyman, J.~D., Lamb, G.~P., Levan, A.~J., et al.\ 2018, arXiv:1801.02669

\bibitem[Meegan et al.(2009)]{meegan09} Meegan, C., Lichti, G., Bhat, P.~N., et al.\ 2009, \apj, 702, 791 

\bibitem[Mooley et al.(2018)]{mooley18} Mooley, K.~P., Nakar, E., Hotokezaka, K., et al.\ 2018, \nat, 554, 207

\bibitem[Piran (1992)]{piran92} \ 1992, \apjl, 389,L45

\bibitem[Planck Collaboration XIII (2016)]{planck} Planck Collaboration, Ade, P.~A.~R., Aghanim, N., et al.\ 2016, \aap, 594, A13 

\bibitem[Preece et al.(2000)]{preece00} Preece, R.~D., Briggs, M.~S., Mallozzi, R.~S., et al.\ 2000, \apjs, 126, 19

\bibitem[Punturo et al.(2010)]{et17}{Punturo}, M., {Abernathy}, M., {Acernese}, F., et al. \ 2010, Classical and Quantum Gravity, 27,194002

\bibitem[Rossi et al.(2002)]{rossi02} Rossi, E., Lazzati, D., \& Rees, M.~J.\ 2002, \mnras, 332, 945

\bibitem[Salafia et al.(2019)]{salafia19} Salafia, O.~S., Ghirlanda, G., Ascenzi, S., et al.\ 2019, arXiv e-prints, arXiv:1905.01190

\bibitem[Sun et al.(2015)]{sun15} Sun, H., Zhang, B., \& Li, Z.\ 2015, \apj, 812, 33 

\bibitem[LIGO Scientific Collaboration (2016)]{ligo16}LIGO Scientific Collaboration  \ 2016, The LSC-Virgo White Paper on Instrument Science (2016-2017 edition)

\bibitem[The LIGO and the Virgo Collaboration (2019a)]{LVC25z} The LIGO and the Virgo Collaboration\ 2019, GRB Coordinates Network, Circular Service, No.~24237

\bibitem[The LIGO and the Virgo Collaboration (2019b)]{LVC26c} The LIGO and the Virgo Collaboration\ 2019, GRB Coordinates Network, Circular Service, No.~24168

\bibitem[Troja et al.(2017)]{troja17} Troja, E., Piro, L., van Eerten, H., et al. 2017, Nature, 551,71T

\bibitem[Troja et al.(2018)]{troja18}{Troja}, E., {Piro}, L., {Ryan}, G., et al.\ 2018, \mnras, 487, L18 

\bibitem[Virgili et al.(2011)]{virgili11}{Virgili}, F.~J., {Zhang}, B., {O'Brien}, P.\& {Troja}, E., \ 2011, \apj, 727,109

\bibitem[Wanderman \& Piran (2015)]{wanderman15}{Wanderman}, D. \& {Piran}, T. \ 2015, \mnras, 448,3026

\bibitem[Xie et al.(2018)]{xie18}{Xie}, X., {Zrake}, J., \& {MacFadyen}, A..\ 2018, \apj, 863, 58 

\bibitem[Yonetoku et al.(2004)]{yonetoku04} Yonetoku, D., Murakami, T., Nakamura, T., et al.\ 2004, \apj, 609, 935

\bibitem[Yuan et al.(2018)]{yuan18} Yuan, W., Zhang, C., Ling, Z., et al.\ 2018, Space Telescopes and Instrumentation 2018: Ultraviolet to Gamma Ray, 1069925

\bibitem[Yuan et al.(2018)]{EP} Yuan, W., Zhang, C., Chen, Y., et al.\ 2018, Scientia Sinica Physica, Mechanica \& Astronomica, 48, 039502

\bibitem[Yuksel et al.(2008)]{yuksel08}{Yuksel}, H., {Kistler}, M.~D., {Beacom}, J.~F., et al. \ 2008, \apjl, 683,L5

\bibitem[Zhang \& M{\'e}sz{\'a}ros(2002)]{zhang02} Zhang, B., \& M{\'e}sz{\'a}ros, P.\ 2002, \apj, 571, 876 

\bibitem[Zhang et al.(2018)]{zhang18} Zhang, B.-B., Zhang, B., Sun, H., et al.\ 2018, Nature Communications, 9, 447 

\bibitem[Zhang et al.(2018)]{gecam} Zhang, D.-L., Li, X.-Q., Xiong, S.-L., et al.\ 2018, arXiv:1804.04499 











\end{thebibliography}
\end{document}